\documentclass{DISproc}

\begin{document}
\title{Z' bosons at the LHC
in a modified MSSM\footnote{Talk given at
DIS2012, XX International Workshop on Deep Inelastic
Scattering and Related Subjects, 26-30 March 2012, Bonn,
Germany.}}

\author{{\slshape G. Corcella}\\[1ex]
INFN, Laboratori Nazionali di Frascati,\\
Via E.~Fermi 40, I-00044, Italy}



\maketitle

\begin{abstract}
I study the production of heavy neutral gauge bosons $Z'$ at the
LHC in U(1)$'$ models, inspired by Grand Unification Theories,
as well as in the Sequential Standard Model, accounting for possible
decays into supersymmetric channels.
I shall consider the MSSM and present results on 
branching ratios and event rates with sparticle
production at the LHC, taking particular care
about final states with charged leptons and missing energy.
\end{abstract}

Heavy neutral bosons $Z'$ are predicted in U(1)$'$ gauge groups,
inspired by Grand Unification Theories (GUTs), and in the
Sequential Standard Model (SSM), wherein the $Z'$ has the same coupling to
fermions and gauge bosons as the $Z$ in the Standard Model (SM)
(see, e.g., the reviews in Refs.~\cite{rizzo,langacker}). From the experimental viewpoint, 
searches for $Z'$ bosons have been performed at the Tevatron \cite{tevatron}
and at the LHC \cite{lhc}. The latest LHC analyses excluded a 
U(1)$'$-based $Z'$ with mass below 2.32 TeV (CMS) and 2.21 TeV (ATLAS),
whereas the lower mass limits for the  SSM $Z'$ are 
currently about 1.49-1.69 TeV (CMS) and 1.77-1.96 TeV (ATLAS).
Such results crucially rely on the assumption that the new neutral
gauge bosons only decay into Standard Model final states.
In this talk, following the lines of \cite{corge}, I wish to investigate
the possibility that the $Z'$ can decay according to modes Beyond the
Standard Model (BSM). In particular, I will consider supersymmetry
and the Minimal Supersymmetric Standard Model, extending the
work carried out in \cite{gherghetta} by varying the $Z'$ mass and
scanning thoroughly the U(1)$'$ and MSSM parameter space.
With respect to Refs.~\cite{baum,chang}, which also
studied $Z'$ decays in supersymmetry,
the so-called D-term correction, due to the extra U(1)$'$ group, is
added to the sfermion masses and the supersymmetric particle masses 
are not treated as free parameters, but they are obtained diagonalizing the
corresponding mass matrices.
If the branching ratios into BSM modes were to be relevant,
one may have to reconsider the current exclusion limits on
the $Z'$ boson. From the point of view of supersymmetry,
the production of charged-slepton or chargino pairs in $Z'$ decays
has the advantage, with respect to other production channels,
that the $Z'$ mass sets a kinematic constrain on the final-state
invariant mass.

The U(1)$'$ models originate from the breaking of a rank-6
GUT group E$_6$ according to ${\rm E}_6\to {\rm SO}(10)\times
{\rm U}(1)'_\psi$, followed by ${\rm SO} (10)\to {\rm SU}(5)
\times {\rm U}(1)'_\chi$.
The heavy neutral bosons associated with ${\rm U}(1)'_\psi$ and 
${\rm U}(1)'_\chi$ are thus named $Z'_\psi$ and $Z'_\chi$, respectively,
whereas a generic $Z'$ boson is a combination of $Z'_\psi$ and $Z'_\chi$,
with a mixing angle $\theta$:
\begin{equation}
Z'(\theta)=Z'_\psi\cos\theta-Z'_\chi\sin\theta.
\end{equation}
Following \cite{corge}, the models and the $Z'$ bosons which will be
investigated are listed in Table~\ref{tabmod}.
The model $Z'_\eta$ comes from the breaking of the GUT group in
the SM, i.e. ${\rm E}_6\to {\rm SM} 
\times {\rm U}(1)'_\eta$;
the $Z'_{\rm S}$ is present in the secluded model, wherein the SM is extended by means of
a singlet field $S$; the $Z'_{\rm N}$ is equivalent to the $Z'_\chi$ 
model,
but with the `unconventional' assignment of SM, MSSM 
and exotic fields 
in the SU(5) representations, as debated in \cite{nardi}.
\begin{wraptable}{r}{0.45\textwidth}
  \centering
  \begin{tabular}{c|c}
  \toprule
    Model & $\theta$ \\
    \midrule
$Z'_\psi$ & 0\\
$Z'_\chi$ & $-\pi/2$\\
$Z'_\eta$ & $\arccos\sqrt{5/8}$\\ 
$Z'_{\rm S}$ & $\arctan(\sqrt{15}/9)-\pi/2$\\
$Z'_{\rm I}$ & $\arccos\sqrt{5/8}-\pi/2$\\
$Z'_{\rm N}$ & $\arctan\sqrt{15}-\pi/2$\\
    \bottomrule
  \end{tabular}
  \caption{$Z'$ bosons in the U(1)$'$ models along with the 
mixing angles.  \label{tabmod}}
\end{wraptable}
When studying supersymmetric contributions to $Z'$ decays, it is necessary
to modify the particle content of the MSSM. First, besides the
MSSM Higgs doublets, one needs an extra
scalar  Higgs boson to break the U(1)$'$ gauge symmetry and give mass to the
$Z'$. After symmetry breaking, one is left with two charged Higgs bosons
$H^\pm$, one neutral CP-odd $A$ and three neutral CP-even, i.e. $h$ and
$H$, already present in the MSSM, and a novel $H'$.
In the gaugino sector, two new neutralinos are to be included, for a total of
six, corresponding
to the supersymmetric partners of the $Z'$ and of the extra Higgs.
However, as pointed out in \cite{corge}, the novel 
Higgs and neutralinos are
typically too heavy to contribute to $Z'$ phenomenology.

In the extended MSSM, besides the SM modes, one has to
consider $Z'$ decays into slepton, squark, chargino, neutralino 
and Higgs pairs,
as well as final states with Higgs bosons associated with a $W$ or a $Z$.
As in \cite{corge}, I shall pay
special attention to supersymmetric decays
yielding  charged leptons, 
as they are the golden channel for the experimental
searches. Final states with two charged leptons and missing energy
may come from primary decays into charged sleptons 
$Z'\to\tilde\ell^+\tilde\ell^-$, with the sleptons decaying 
according to $\tilde\ell^\pm\to \ell^\pm\tilde\chi^0$, 
with $\tilde\chi^0$ being a neutralino, or from primary decays into
charginos $Z'\to\tilde\chi_2^+\tilde\chi_2^-$, 
followed by $\tilde\chi_2^\pm\to \ell^\pm\tilde\chi_1^0$.
A decay chain, leading to four leptons and missing energy, 
is also yielded by decays into neutralinos $Z'\to\tilde\chi_2^0\chi_2^0$, with
subsequent $\tilde\chi_2^0\to \ell^\pm\tilde\ell^\mp$ and 
$\tilde\ell^\pm\to\ell^\pm\tilde\chi_1^0$.
Finally, the decay into sneutrino pairs, such as
$Z'\to\tilde\nu_2\tilde\nu_2^*$, followed by  $\tilde\nu_2\to  \tilde\chi^0_2\nu$
and $\tilde\chi_2^0\to \ell^+\ell^-\tilde\chi^0_1$, 
with an intermediate charged slepton, gives four
charged leptons and missing energy, due to both neutrinos and neutralinos.

Following \cite{corge}, I explore the $Z'$ branching ratios by varying the
$Z'$ and slepton masses, while 
fixing the other parameters to the
following `Reference Point':
\begin{eqnarray}
&\ &
\mu =200\ ,\ \tan\beta=20\ ,\ A_q=A_\ell=A_f=500~{\rm GeV}\ ,\nonumber \\
&\ &m^0_{\tilde q}=5~{\rm TeV}\ ,\ 
 M_1=150~{\rm GeV}\ ,\ M_2=300~{\rm GeV}\ ,\ M^\prime=1~{\rm TeV}.
\label{refpoint}
\end{eqnarray}
In Eq.~(\ref{refpoint}), $\mu$ is the parameter contained in the 
Higgs superpotential, $\tan\beta=v_2/v_1$ is the ratio of the vacuum
expectation values of the two MSSM Higgs doublets, $A_f$ is the coupling
of the Higgs with the fermions.
Furthermore, $m^0_{\tilde q}$ is the squark mass, assumed
to be the same for all flavours at the $Z'$ scale, before the
addition of the D-term, $M_1$, $M_2$ and $M'$ are the soft masses of
the gauginos $\tilde B$, $\tilde W_3$ and $\tilde B '$.
As for the U(1)$'$ coupling $g'$, we shall adopt the GUT-driven convention
that it is proportional to the coupling constant
$g_1$ of U(1) via $g'=\sqrt{5/3}g_1$.
In the Sequential Standard Model, the coupling of the $Z'_{\rm SSM}$ to sfermions
is instead
the same as in the SM, i.e. $g_{\rm{SSM}}=g_2/(2\cos\theta_W)$,
where $g_2$ is the SU(2) coupling and $\theta_W$ the Weinberg angle.

An extensive analysis for all the models quoted in Table~\ref{tabmod} 
has been carried out in \cite{corge}: here I just report the branching ratios
for the GUT-inspired model $Z'_\psi$ (Table~\ref{psi})
and for the $Z'_{\rm SSM}$ (Table~\ref{ssm}), since they are the ones yielding
the highest rates into supersymmetric channels.
The branching ratios are
listed for 1~TeV$<m_{Z'}<$5~TeV and for the values of $m^0_{\tilde\ell}$ which
minimize and maximize the slepton rate,  with $m^0_{\tilde\ell}$ being the
slepton mass, assumed to be the same for all flavours, before the D-term addition.
In the $Z'_\psi$ case, both SM and BSM branching fractions are 
reported; 
for the SSM, only the BSM
channels are quoted.
From such tables, one can learn that, in the $Z'_\psi$ scenario, the BSM
modes account for about 35-40\% of the total width, whereas in the SSM
they can be up to 60-65\%. In both cases, the dominant BSM contributions
are the ones into neutralinos and charginos, whereas the slepton modes,
i.e. charged sleptons or sneutrinos, can reach 4\% for the
$Z'_\psi$ and 5-6\% for the $Z'_{\rm SSM}$.  Ref.~\cite{corge}
also presents the branching ratios as a function of $m^0_{\ell}$:
as expected, the slepton rates rapidly decrease as $m^0_{\tilde\ell}$
increases. Such spectra are not shown here for the sake of brevity.
\begin{table}
  \centering
  \begin{tabular}{cccccccccccc}
  \toprule
$m_{Z'}$ & $m^0_{\tilde\ell} $   & BR$_{q\bar q}$ &BR$_{\ell\ell}$  & 
BR$_{\nu\bar \nu}$ & BR$_{WW}$          &    BR$_{Zh}$
& BR$_{\tilde\chi^+\tilde\chi^-}$ &BR$_{\tilde\chi^0\tilde\chi^0}$  
&  BR$_{\tilde\nu\tilde\nu^*}$  & BR$_{\tilde\ell\tilde\ell}$ &  
$\rm{BR}_{\rm{BSM}}$\\
    \midrule
1.0    & 0.4 & 48.16 & 8.26 & 8.26 & 3.00 & 2.89 & 9.13 & 16.53 & 
1.91 & 1.90 & 35.31 \\
1.0    & 0.7	& 50.07 & 8.59 & 8.59 & 3.08 & 2.99 & 9.49 & 17.18 & 0.00 & 
0.00 & 32.75 \\ 
1.5    & 0.6	& 46.78 & 7.90 & 7.90 & 2.71 & 2.69 & 9.73 & 18.64 & 
1.83 & 1.83 & 37.43\\
1.5    & 1.0 & 48.55 & 8.20 & 8.20 & 2.81 & 2.79 & 10.10 & 19.35 & 0.00 & 
0.00 & 35.05 \\
2.0   & 0.8 & 46.30 & 7.77 & 7.77 & 2.62 & 2.62 & 9.92 & 19.37 & 
1.80 & 1.80 & 38.15\\
2.0   &1.3 & 48.03 & 8.06 & 8.06 & 2.72 & 2.72 & 10.29 & 20.10 & 0.00 & 
0.00 & 35.84 \\
2.5   &1.0 & 46.01 & 7.70 & 7.70 & 2.58 & 2.59 & 9.99 & 19.68 &
1.79 & 1.78  & 38.58 \\
2.5   &1.6	& 47.72 & 7.99 & 7.99 & 2.67 & 2.68 & 10.36 & 20.41 & 
0.00 &    0.00 & 36.30 \\
3.0 & 1.1 & 45.35 & 7.58 & 7.58 & 2.53 & 2.54 & 9.92 & 19.63 & 1.86 
& 1.86 & 39.49 \\
3.0 & 1.9 & 47.10 & 7.88 & 7.88 & 2.62 & 2.64 & 10.30 & 20.39 & 0.00 & 
0.00 & 37.15 \\
3.5 & 1.3 & 44.91 & 7.50 & 7.50 & 2.49 & 2.51 & 9.86 & 19.58 & 
1.83 & 1.83 & 40.08	\\	
3.5 & 2.2 & 46.61 & 7.79 & 7.79 & 2.59 & 2.61 & 10.24 & 20.32 & 0.00 & 
0.00 & 37.81\\
4.0 & 1.5 & 44.60 & 7.45 & 7.45 & 2.47 & 2.49 & 9.82 & 19.53 & 
1.80 & 1.80 & 40.51\\
4.0 & 2.5 & 46.26 & 7.72 & 7.72 & 2.56 & 2.58 & 10.19 & 20.26 & 0.00 & 
0.00 &    38.29 \\
4.5 & 1.6 & 44.32 & 7.40 & 7.40 & 2.45 & 2.47 & 9.78 & 19.47 & 
1.84 & 1.84 & 40.89\\
4.5 & 2.8 & 46.01 & 7.68 & 7.68 & 2.54 & 2.57 & 10.15 & 20.21 & 0.00 & 0.00 & 
 38.63\\
5.0 & 1.8 & 44.16 & 7.37 & 7.37 & 2.44 & 2.46 & 9.76 & 19.44 & 
1.82 & 1.82 &  41.11\\
5.0 & 3.1 & 45.83 & 7.65 & 7.65 & 2.53 & 2.55 & 10.13 & 20.18 & 0.00 & 
0.00 & 38.88\\
    \bottomrule
  \end{tabular}
  \caption{$Z'_\psi$ branching ratios for a few
values of $Z'$ and slepton masses, expressed in TeV.
BR$_{\rm BSM}$ is the total decay rate in BSM channels. \label{psi}}  
\end{table}\par
Before concluding, in Table~\ref{number} I present the expected number of
events with supersymmetric cascades ($N_{\rm casc}$), 
i.e. production of
neutralinos, charginos or sleptons, 
and the charged-slepton rates ($N_{\rm slep}$),
in the high-luminosity phase of the LHC, i.e. ${\cal L}=100~{\rm fb}^{-1}$,
and at the centre-of-mass energy $\sqrt{s}=14$~TeV. 
The parameters are fixed to
the Reference Point (\ref{refpoint}), whereas the $Z'$ is set either to 1.5 or
to 2 TeV and $m^0_{\tilde\ell}$ to 
the value maximizing the slepton
rate. The numbers in Table~\ref{number} are obtained in the
narrow-width approximation and calculating the $pp\to Z'$ cross section
at leading order, as in \cite{corge}.
One finds that the cascade events can be up to 
${\cal O}(10^5)$ and the charged sleptons
up to ${\cal O}(10^4)$: the highest rate of production
of supersymmetric particles occurs in the SSM,
but even the U(1)$'$ models yield meaningful sparticle production.

In summary, I reviewed the main issues discussed in Ref.~\cite{corge}
and presented some results on $Z'$ decays in the MSSM,
extended by means of an extra GUT-inspired
U(1)$'$ group, as well as in the SSM.
In order to reconsider the $Z'$ exclusion limits or
draw a conclusive statement on the feasibility to discover 
supersymmetry in $Z'$ decays at the LHC, however, it will be compulsory
implementing this modelling in the framework of a Monte Carlo
generator. This is in progress.

\begin{table}[htbp]
  \centering
  \begin{tabular}{cccccccccc}
  \toprule
$m_{Z'}$ & $m^0_{\tilde\ell}$   &
BR$_{H^+H^-}$ & BR$_{Zh}$ & B$_{hA}$ 
& BR$_{\tilde\chi^+\tilde\chi^-}$ & BR$_{\tilde\chi^0\tilde\chi^0}$ 
& BR$_{\tilde\ell\tilde\ell}$ & BR$_{\tilde\nu\tilde\nu^*}$ & BR$_{\rm{BSM}}$		\\
    \midrule
1.0 & 0.10 & 0.00 & $\sim 10^{-6}$ &  0.00 & 18.31 & 
29.30 & 1.89 & 3.77 &  53.27\\	 
1.0 & 0.50 & 0.00 & $\sim 10^{-6}$ & 0.00 & 19.41 &  
31.06 & 0.00 & 0.00 & 50.47 \\
1.5 & 0.10 & 0.00 & 0.87 & 0.76 & 17.84 &
32.52 & 1.75 & 3.48 & 57.21 \\
1.5 & 0.75 & 0.00 & 0.92 & 0.80 & 18.82 &
34.31 & 0.00 & 0.00 & 54.55\\
2.0 & 0.10 & 0.00 & 1.93 & 1.85 & 17.37 &  
33.01 & 1.67 & 3.33 &  59.17\\
2.0 & 1.00 & 0.00 & 2.04 & 1.95 & 18.28 & 
34.75 & 0.00 & 0.00 & 57.02 \\
2.5 & 0.10 & 0.91 & 2.59 & 2.53 &  
16.93 & 32.78 & 1.62 & 3.22 & 60.58\\
2.5 & 1.25 & 0.95 & 2.72 & 2.66 & 
17.79 & 34.45 & 0.00 & 0.00 & 58.57\\
3.0 & 0.10 & 1.72 & 2.98 & 2.94 &  
16.62 & 32.51 & 1.58 & 3.15 & 61.49 \\
3.0 & 1.50 & 1.81 & 3.13 & 3.08 &  
17.44 & 34.12 & 0.00 & 0.00 & 59.58 \\
3.5 & 0.10 & 2.27 & 3.23 & 3.20 &  
16.42 & 32.30 & 1.56 & 3.10 & 62.08 \\
3.5 & 1.75 & 2.38 & 3.38 & 3.35 &  
17.22 & 33.88 & 0.00 & 0.00 & 60.22 \\
4.0 & 0.10 & 2.65 & 3.39 & 3.37 &  
16.28 & 32.16 & 1.54 & 3.07 & 62.46\\
4.0 & 2.00 & 2.78 & 3.56 & 3.53 & 
17.07 & 33.71 & 0.00 & 0.00 & 60.65 \\
4.5 & 0.10  & 2.91 & 3.51 & 3.49 &  
16.19 & 32.06 & 1.53 & 3.05 & 62.73 \\
4.5 & 2.25 & 3.05 & 3.67 & 3.65 &  
16.96 & 33.59 & 0.00 & 0.00 & 60.94 \\
5.0 & 0.10 & 3.11 & 3.59 & 3.57 &
16.12 & 31.98 & 1.52 & 3.03 & 62.93\\
5.0 & 2.50 & 3.26 & 3.76 & 3.74 & 
16.89 & 33.51 & 0.00 & 0.00 & 61.16\\
    \bottomrule
  \end{tabular}
  \caption{As in Table~\ref{psi}, 
but for the Sequential Standard Model, 
including only the BSM modes.}
  \label{ssm}
\end{table}
\begin{wraptable}{r}{0.45\textwidth}
\centering
  \begin{tabular}{cccccc}
  \toprule
Model & $m_{Z'}$ & N$_{\rm casc}$ & N$_{\rm slep}$ \\ 
 \midrule
$Z'_\eta$ & 1.5  &  13650 & -- \\
$Z'_\eta$ & 2.0 &  2344 & -- \\
$Z'_\psi$ & 1.5 &  10241 & 622 \\
$Z'_\psi$ & 2.0 & 2784 & 162 \\
$Z'_{\rm N}$ & 1.5 &  9979 & 414 \\
$Z'_{\rm N}$ & 2.0 &  2705 & 104 \\
$Z'_{\rm I}$ & 1.5 &  8507 & -- \\
$Z'_{\rm I}$ & 2.0 & 2230 & -- \\
$Z'_{\rm S}$ & 1.5 & 8242 & 65 \\
$Z'_{\rm S}$ & 2.0 & 2146  & 16 \\
$Z'_{\rm SSM}$ & 1.5 &  775715  & 24774 \\
$Z'_{\rm SSM}$ & 2 & 19570 & 606 \\
    \bottomrule
  \end{tabular}
  \caption{Rates of supersymmetric cascades and charged sleptons at the LHC
for an integrated luminosity of 100 fb$^{-1}$ and a centre-of-mass energy
of 14 TeV. The $Z'$ mass is given in TeV.}
  \label{number}
\end{wraptable}

\begin{footnotesize}

\end{footnotesize}

\end{document}